\documentclass[]{iaus}
\usepackage{graphicx,bm, natbib}

\title[Discs, outflows, and feedback] %% give here short title %%
{Discs, outflows, and feedback in
  collapsing magnetized cores}

\author[Dennis F. Duffin \& Ralph E. Pudritz]   %% give here short author list %%
{Dennis F. Duffin$^1$
 \and Ralph E. Pudritz$^2$}

\affiliation{$^1$Department of Physics and Astronomy, McMaster University, \\ Hamilton ON, L8S 4M1, Canada \\ email: {\tt duffindf@mcmaster.ca} \\[\affilskip]
$^2$Origins Institue, McMaster University \\ Hamilton ON, L8S 4M1, Canada \\ email: {\tt pudritz@physics.mcmaster.ca}}

\pubyear{2010}
\volume{270}  %% insert here IAU Symposium No.
\pagerange{1--4}
\jname{Computation Star Formation}
\editors{J. Alves, B. Elmegreen, J.M. Girart \& V. Trimble, eds.}
\begin{document}

\maketitle

\begin{abstract}
The pre-stellar cores in which low mass stars form are generally well magnetized.  Our simulations show that early protostellar discs are massive and experience strong magnetic torques in the form of magnetic braking and protostellar outflows.  Simulations of protostellar disk formation suggest that these torques are strong enough to suppress a rotationally supported structure from forming for near critical values of mass-to-flux.  We demonstrate through the use of a 3D adaptive mesh refinement code -- including cooling, sink particles and magnetic fields -- that one produces transient 1000 AU discs while simultaneously generating large outflows which leave the core region, carrying away mass and angular momentum.  Early inflow/outflow rates suggest that only a small fraction of the mass is lost in the initial magnetic tower/jet event.
\end{abstract}

\firstsection % if your document starts with a section,
\section{Introduction}

The early evolution of low mass, isolated protostellar cores in the pre-Class 0 stage is now understood as the interplay between gravity, rotation, magnetic fields, and radiative cooling. Strong magnetic fluxes have been observed in molecular clouds \citep[e.g.][]{1999ApJ...520..706C}. Recent simulations have focused on the changes endued by a magnetic field in the collapse, through either a 3D SPHR or AMR approach \citep{1994ApJ...432..720B,2004MNRAS.348L...1M, HW2004b,  2005MNRAS.362..369M,2005MNRAS.362..382M, 2006ApJ...641..949B, 2007MNRAS.377...77P, 2008A&A...477....9H, 2009ApJ...706L..46D}.  It has been shown that magnetic fields slow the collapse timescale and brake the rotation of the initial core and of massive disc-like structures that subsequently form. Magnetic fields have also been shown to  suppress fragmentation and the formation of bars and spiral waves.  Furthermore, they facilitate the launching of molecular outflows.

Several theoretical problems have arisen from these simulations.  First, from axisymmetric simulations outside of 6.7 AU, it is argued that Keplerian discs cannot form in an ideal MHD collapse \citep{2008ApJ...681.1356M}, except in the limit of very weak magnetic flux or high magnetic diffusivity whether numerical \citep{2010ApJ...716.1541K} or through strong ambipolar diffusion \citep{2009ApJ...698..922M}. Secondly, magnetic tension and pressure seem to efficiently suppress fragmentation in the early stages of protostellar collapse so that it is difficult to form multi-star systems.
With the ability to extend the magnetized collapse further, we can begin to examine questions such as how the Core Mass Function (CMF) is related to the Initial Mass Function of stars \citep[e.g.][]{2000ApJ...545..364M}, the nature of the molecular outflow on larger scales and whether fragmentation is suppressed even in later stages. 
We present our results of the early collapse using 3D ideal and non-ideal magnetohydrodynamic simulations, and the first results of the later stages of the magnetized collapse and outflow.  We are able to evolve the simulation an additional $10^4-10^5$ yr due to the implementation of the sink particle in the FLASH AMR code \citep{2010ApJ...713..269F}. %Rotationally dominated structures form early on, while radial velocities are still large. This structure grows with time.  In the limit of no magnetic field, we show that bar modes strongly brake the massive disk, preventing a rotationally domitated structure from forming. We simulate our collapse using ambipolar diffusion to demonstrate a possible resolution the magnetic fragmentation crisis.  Finally, we discuss the possible connections between the CMF and IMF.  
  
\section{Numerical Methods and Initial Conditions}

We model our a stellar core as in previous papers \citep[e.g.][]{2009ApJ...706L..46D, 2004MNRAS.355..248B}, by embedding a slightly over-critical 1 $M_\odot$ Bonnor-Ebert sphere \citep{1956MNRAS.116..351B,1955ZA.....37..217E} in a low density environment.  We add to this density distribution a 10\% over-density to ensure collapse and a 10\% $m=2$ perturbation to break symmetry \citep[see e.g.][]{2009ApJ...706L..46D,2004MNRAS.355..248B}. The background is an isothermal, low density environment in pressure equilibrium with the sphere (the density is set by choosing a background that is 10 times warmer than the sphere).  The box is roughly 10 times the size of the BE radius (0.81 pc in these models). We add to the sphere uniform rotation such that the ratio of rotational and gravitational energy is moderate ($\beta_\mathrm{rot}$=0.046, similar to the value used in \citet{2008A&A...477....9H}) and a constant $\beta_\mathrm{plasma} = 2c_s^2/v_\mathrm{A}^2  = 46.01$.  In this model, one can relate the mass to flux ratio $\mu/\mu_0$, where $\mu_0$ is the critical mass to flux of $\mu_0 = (2\pi G^{1/2})^{-1}$,  by $\mu/\mu_0 \simeq 0.74 c_s / v_\mathrm{A}$ = 3.5.

We use sink radii of 12.7 AU for longer runs, and radii of 3.2 AU to test the effect of sink particle size on the result.  The accretion radius of a sink corresponds to 2.5 cells at the highest refinement level, and the critical gas density (beyond which gas can be accreted into particles) corresponds to the Jeans' density at the core temperature (20 K) of these cells.  This gives $\rho_\mathrm{acc} = 3.69\times 10^{-12}~\mathrm{g~cm^{-3}}$ and $\rho_\mathrm{acc} = 5.91\times 10^{-11}~\mathrm{g~cm^{-3}}$ for 12.7 and 3.2 AU sinks respectively. Each Jeans' length is refined by at least 8 cells and de-refined for more than 32 cells. Our customized version of the FLASH AMR code is described in previous work \citep{2000ApJS..131..273F, 2006MNRAS.373.1091B, 2008MNRAS.391.1659D,2010ApJ...713..269F}.

\section{The Early Collapse Phase}

The advantage of \emph{not} using a sink particle is that we properly resolve the gas in the collapse. The disadvantage is that we are limited to pre-Class 0 times ($\approx 10^5$ yr). We compare the following early collapse models: i) ambipolar diffusion, ii) ideal MHD and iii) hydrodynamics (e.g.~$\bm{B}=0$). 
%These results had earlier been used to argue the formation of early accretion disk structures (10 AU in size, a few 10$^5$ yr into the collapse) in the collapse \citep{2009ApJ...706L..46D} that coincided with the launching of early protostellar outflows.  These smooth disk structures only persisted in the magnetic collapses which surpressed the strong bar modes present in the hydrodynamic collapse. 
The rotational properties of these early structures are shown in Figure 1a.  We compare surface density averaged values of $v_\phi/v_r$ and $v_\phi/v_\mathrm{Kepler}$ to measure the degree of rotation in these early collapsed structures. The comparison is done at the maximum common central surface density ($\Sigma_z$).  This is limited primarily by the ambipolar diffusion model which has the most constrained timestep.  More evolved versions of ideal MHD and hydro models are included as thin lines in the graph, hinting at the evolution of the possible structures. The inward velocity has not halted (e.g. the flow is sub-Keplerian)  in these early structures, however we are indeed seeing flattened rotationally dominated accretion discs (with associated outflows) being generated with sizes of 6-7 AU, facilitated by the suppression of gravitational instabilities. Bars produced in the hydrodynamic model have prevented a similar structure from forming by efficiently redistributing angular momentum.

To demonstrate the suppression of fragmentation further, we ran the simulation with 10 times more rotational energy in order to study early fragmentation ($\beta_\mathrm{rot}$ = 0.74, corresponding to high values seen in simulations of core formation in a turbulent medium \citep{2007MNRAS.382...73T}). Indeed our hydrodynamic collapse produces a wide binary with a separation of about 1000 AU. The ideal MHD model suppresses the fragmentation, as documented in previous studies \citep{2007MNRAS.377...77P, 2008A&A...477...25H}, resulting in a large bar and a central condensed structure.  Meanwhile, the ambipolar diffusion model produces an intermediate result, a bar that has fragmented to form a binary.  It perhaps through ambipolar diffusion that any ``fragmentation crisis'' can be solved.  These results are discussed further in \citep{2009ApJ...706L..46D}. 

\begin{figure}
\begin{tabular}{cc}
\includegraphics[width=6cm]{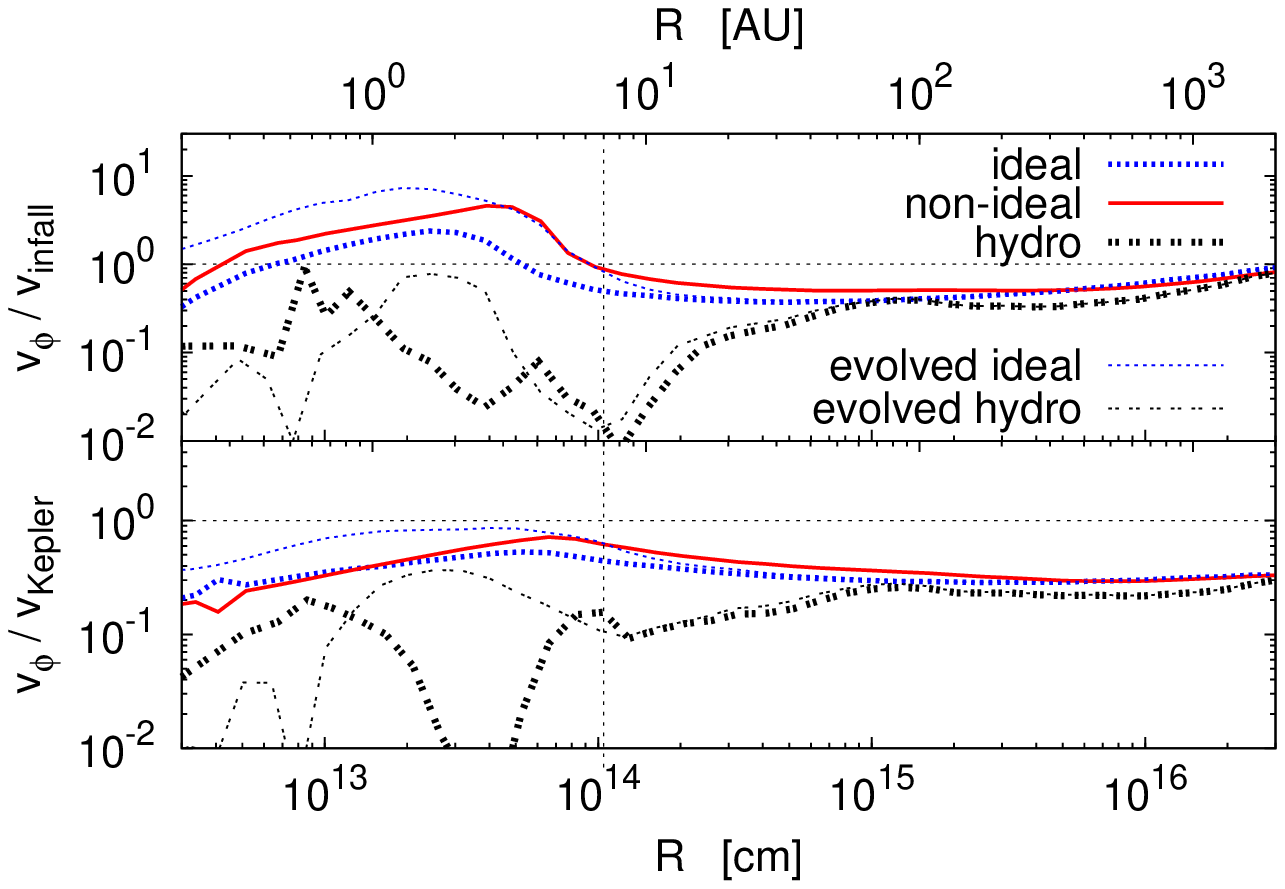} & 
\includegraphics[width=6cm]{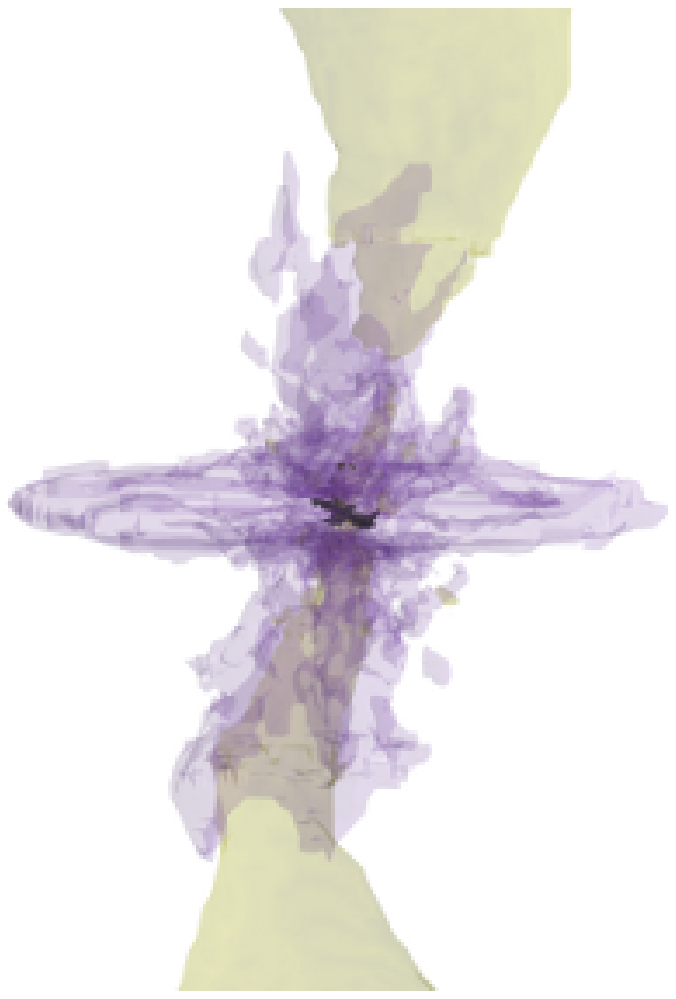} \\
Figure 1a & Figure 1b \\
\includegraphics[width=6cm]{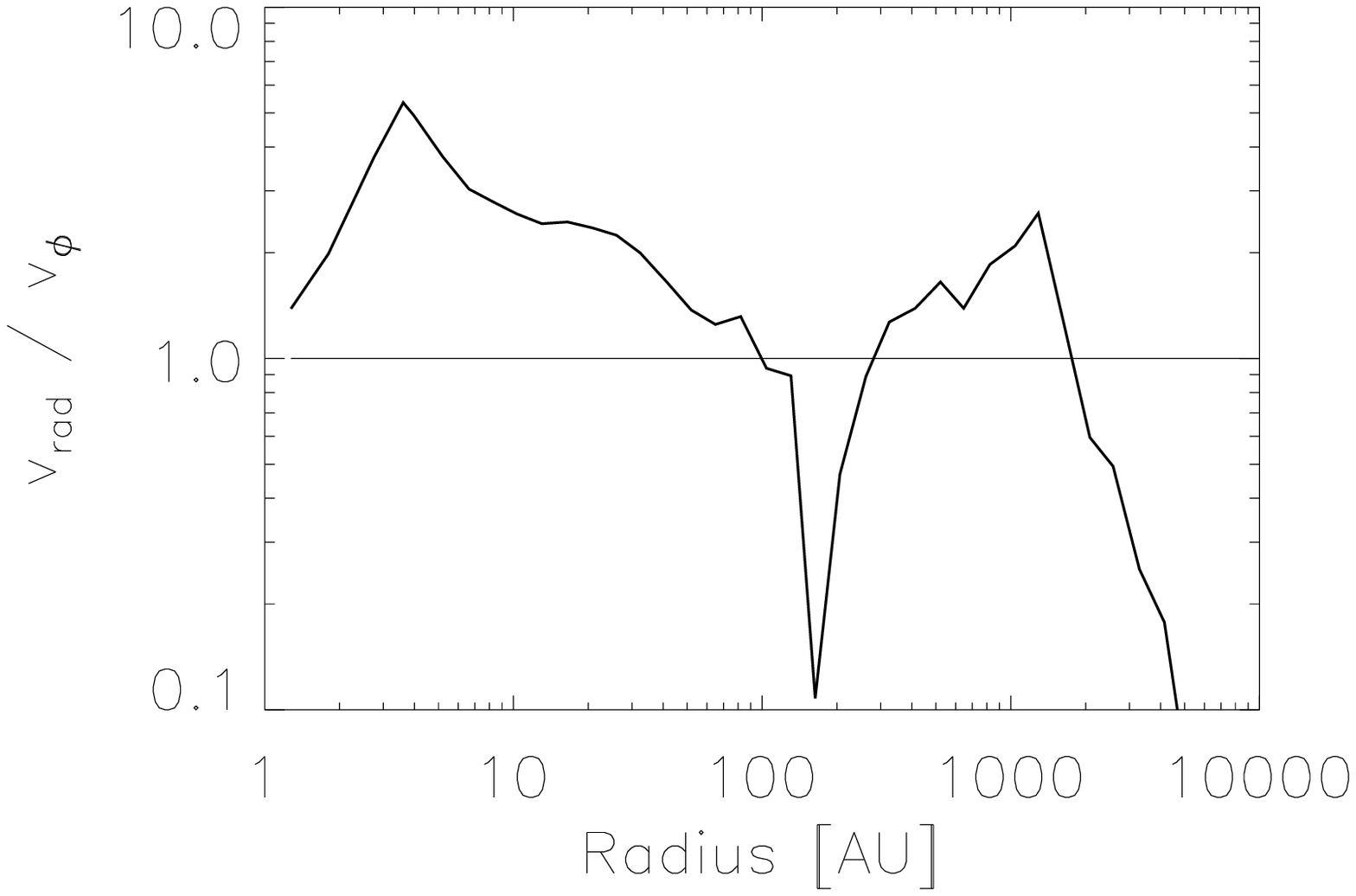} & 
\includegraphics[width=6cm]{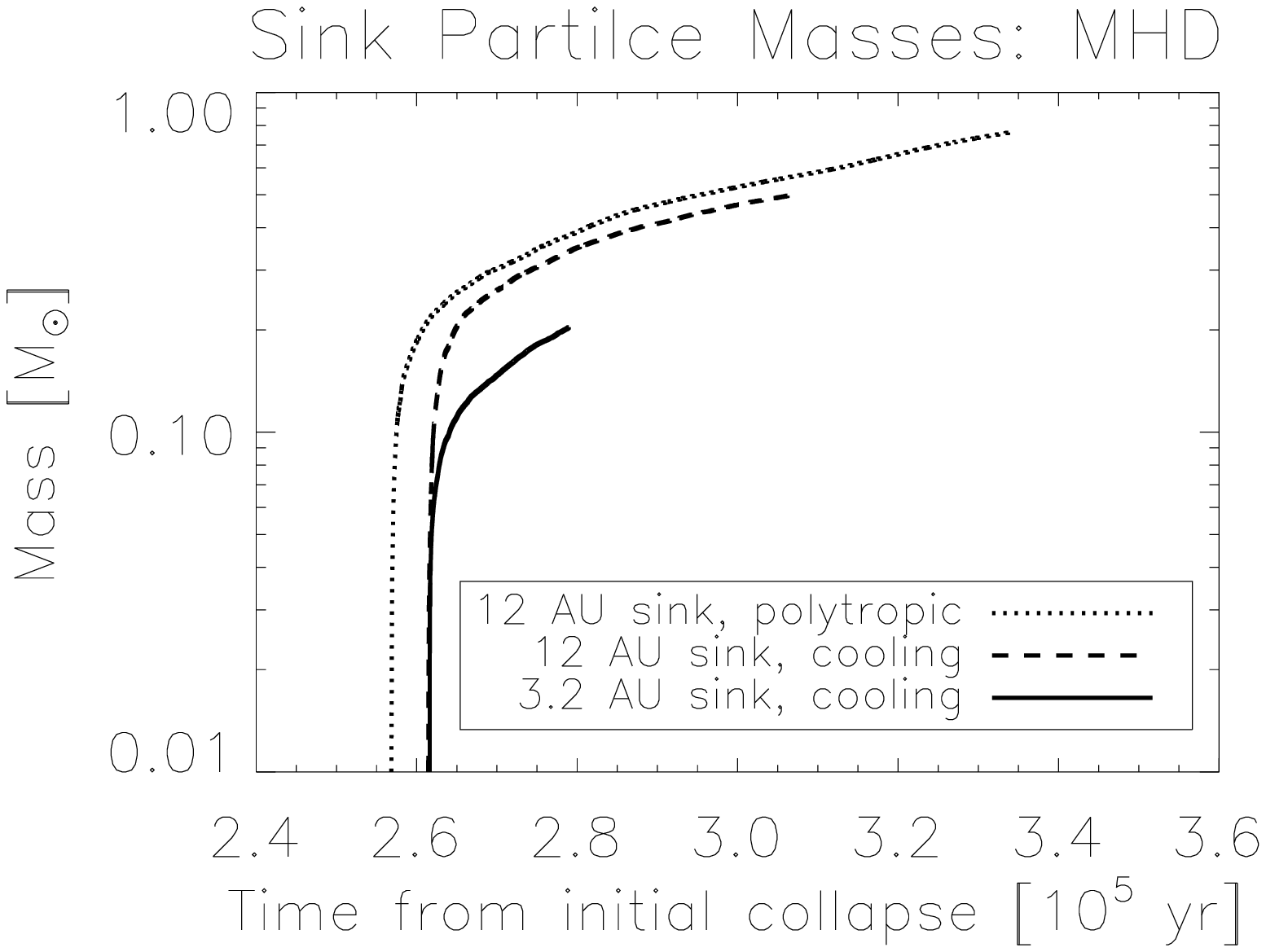} \\
Figure 1c & Figure 1d
\end{tabular}
\caption{\label{fig:1}The early to late evolution of the protostellar core with ideal MHD.  In a). \citep[adapted from][]{2009ApJ...706L..46D}, the rotation properties $v_\phi/v_r$ and $v_\phi/v_\mathrm{Kepler}$ of the early collapse (no sink particle). In b), contours of density (purple or medium grey, $3.3\times10^{-17}~\mathrm{g~cm^{-3}}$ and black, $1.33\times10^{-15}~\mathrm{g~cm^{-3}}$ at 2000 and 100 AU radii respectively) and $v_z$ (yellow or light grey, $1 km/s$) at the end of the simulation (with 3.2 AU sinks).  In c), rotational properties of disk at the end of the simulation (with 3.2 AU sink).  In d), mass evolution of different sink particle models, from left to right:  12 AU sink particle and a polytropic equation of state, 12 AU sink particle and 3.2 AU sink particle both with molecular cooling.
}
\end{figure}

\section{The Later Evolution: Sink Particles}

Using sink particles and ideal MHD, we were able to take the next step and evolve the collapse to much later times. The result is shown in Figure 1b, wherein evolution over an additional 30 kyr is achieved. The sink particle mass evolution is shown in Figure 1d for 12 AU sink particles (with molecular cooling or polytropic equation of state similar to that of \citet{2010arXiv1001.1404M}) and a smaller 3.2 AU sink particle with molecular cooling \citep{2006MNRAS.373.1091B}.  By the end of the smaller sink collapse, the sink particle is $0.2 M_\odot$ and the outflow has grown to $10^4$ AU above the mid-plane of the disk. We are able to run the simulations of larger sinks further, and indeed these particles end up with  nearly all of the core mass (80-90\%).  This is much higher than estimates of theoretical models seeking to relate the Core Mass Function (CMF) to the Initial Mass Function (IMF)  \citep[e.g.][]{2000ApJ...545..364M}. These results suggest that not much mass has in fact been cleared out by the outflow.  This may be due to the fact that most mass must settle into the accretion disk before the outflow is launched, leaving less mass available to be cleared.

The accretion disk is represented by the ratio of $v_\phi/v_r$ in Figure 1c, showing a 2000 AU accretion disk, which appears to be dissipating by the end of the simulation (as seen in Figure 1d). There are two types of outflows, one lower speed ($\lesssim 1$ km/s) and a central, higher speed centrifugally driven wind.  
The inner disk and central component of the outflow are warped and precessing, as shown in Figure 1b. The appearance of precessing warped discs and their outflow is extremely interesting and is related to the back reaction of the MHD outflow on the disk \citep[e.g.][]{2003ApJ...591L.119L}. Many of the qualitative properties of molecular outflows, including jet precession, clumpiness and an onion layered velocity structure \citep{2007prpl.conf..277P} are seen in our simulation, and occur naturally as a consequence of solving the gravito-magnetohydrodynamic equations.

\section{Summary}

In the early stages, accretion discs are small ($< 10$ AU), massive, flattened, rotationally dominated, and are held together by the magnetic field.  As the collapse continues over an additional $10^5$ yr, the accretion disc and its associated outflow grow in size.  The outflow torques and warps the disk leading to disc and jet outflow and precession respectively.  Most material will be accreted by the star raising issues concerning the extent of protostellar feedback on stellar masses.

%\begin{figure}
%\begin{tabular}{cc}
%\includegraphics[width=6cm]{./figs/fig-2a.eps} & 
%\includegraphics[width=6cm]{./figs/fig-3a.eps} \\
%Figure 1a & Figure 1b \\
%\includegraphics[width=6cm]{./figs/fig-3b.eps} & 
%\includegraphics[width=6cm]{./figs/fig-3c.eps} \\
%Figure 1c & Figure 1d
%\end{tabular}
%\caption{The early structures}
%\end{figure}

%\begin{figure}
%\begin{tabular}{cc}
%\includegraphics[width=6cm]{./figs/M1b-bouchut-mhd_hdf5_plt_cnt_2487_white.ps} & 
%\includegraphics[width=6cm]{./figs/mass-time-mhd-models.ps} \\
%Figure 2a & Figure 2b \\
%\includegraphics[width=6cm]{./figs/m1b-bouchut-mhd-2488-vrvp.eps} & 
%\includegraphics[width=6cm]{./figs/m1b-bouchut-mhd-2488-mdot-in-out-ratio.eps} \\
%Figure 2c & Figure 2d
%\end{tabular}
%\caption{The late structures}
%\end{figure}

\bibliographystyle{apj} 
\bibliography{/home/duffindf/school/papers-2008-laptop/master}

\end{document}